\documentclass[conference, 10pt]{IEEEtran}
\IEEEoverridecommandlockouts
\usepackage{cite}
\usepackage{amsmath,amssymb,amsfonts}
\usepackage{algorithmic}
\usepackage{graphicx}
\usepackage{diagbox}
\usepackage{textcomp}
\usepackage{xcolor}
\usepackage{scalerel}
\usepackage{csquotes}
\usepackage{dsfont}
\usepackage{geometry}
\usepackage{subcaption}
\usepackage{float}
\usepackage{amsmath,amsthm}
\usepackage[]{graphicx}

\usepackage{lipsum}
\def\BibTeX{{\rm B\kern-.05em{\sc i\kern-.025em b}\kern-.08em
    T\kern-.1667em\lower.7ex\hbox{E}\kern-.125emX}}

\newgeometry{left=1.57cm,right=1.57cm,top=0.7in,bottom=2.54cm}

\begin{document}

\title{Fog-Based Detection for Random-Access IoT Networks with Per-Measurement Preambles\\
\thanks{This work has received funding from the European  Research  Council (ERC) under the European Union Horizon 2020 research and innovation program (grant agreements 725731 and 648382).}
}

\author{\IEEEauthorblockN{Rahif Kassab\IEEEauthorrefmark{1},
Osvaldo Simeone\IEEEauthorrefmark{1} and Petar Popovski\IEEEauthorrefmark{2} }\\
\IEEEauthorblockA{\small\IEEEauthorrefmark{1}King's Communications, Learning, and Information Processing (KCLIP) Lab, King's College London, London, United Kingdom\\
\IEEEauthorrefmark{2}Department of Electronic Systems, Aalborg University, Aalborg, Denmark\\
Emails: \IEEEauthorrefmark{1}\{rahif.kassab,osvaldo.simeone\}@kcl.ac.uk,
\IEEEauthorrefmark{2}petarp@es.aau.dk}}

\maketitle

\begin{abstract}
Internet of Things (IoT) systems may be deployed to monitor spatially distributed quantities of interests (QoIs), such as noise or pollution levels. This paper considers a fog-based IoT network, in which active IoT devices transmit measurements of the monitored QoIs to the local edge node (EN), while the ENs are connected to a cloud processor via limited-capacity fronthaul links. While the conventional approach uses preambles as metadata for reserving communication resources, here we consider assigning preambles directly to measurement levels across all devices. The resulting Type-Based Multiple Access (TBMA) protocol enables the efficient remote detection of the QoIs, rather than of the individual payloads. The performance of both edge and cloud-based detection or hypothesis testing is evaluated in terms of error exponents. Cloud-based hypothesis testing is shown theoretically and via numerical results to be advantageous when the inter-cell interference power and the fronthaul capacity are sufficiently large.
\end{abstract}


\section{Introduction}
\label{sec:introduction}
The density of connected wireless devices is expected to continue growing as 5G and beyond-5G systems are deployed, especially for Internet-of-Things (IoT) services supported by massive Machine-Type Communications \cite{ding20196g}. This motivates the investigation of access schemes that support high device densities without penalizing the end-to-end performance for specific IoT services.
\begin{figure}[t]
	\centering
	\includegraphics[height= 6 cm, width= 6 cm]{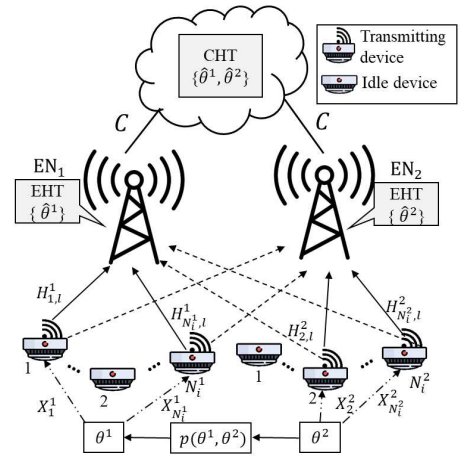}
	\caption{Multi-cell fog-based IoT network aimed at estimating correlated distributed quantities of interest (QoIs) through either edge or cloud processing. }
	\label{fig:system_model} \vspace{-6 mm}
\end{figure}
In this paper, we address this problem by considering the fog-radio access network deployment illustrated in Fig. 1, in which IoT devices monitor distributed Quantities of Interest (QoIs), such as noise or pollution levels. The devices access the network through their local Edge Nodes (ENs), e.g., access points, which are in turn connected via fronthaul links to a cloud processor. Devices are interrogated periodically from the corresponding EN, and they only transmit their measurements of the QoIs if active. The goal of the network is to detect the distributed QoIs based on hypothesis testing at either ENs or at the cloud. 

A conventional approach would prescribe a random access protocol, such as ALOHA, through which devices communicate individual payloads to the local ENs. In case two or more devices select the same preamble during the random access phase, a collision would occur and no information would be delivered to the network. As recognized for \emph{single-cell systems} in \cite{anandkumar2007type,mergen2006tbma, tbma_sayeed}, when the goal is estimating a common QoI measured at multiple devices, the requirement of distinct preambles per active user for a successful transmission is unnecessary. In such a case, it is in fact potentially more efficient to assign a specific preamble to each measurement level: in this way, devices making the same measurement contribute energy to the same preamble, potentially reinforcing its detection signal-to-noise ratio. The estimate of the QoI can then be obtained from the histogram of the received measurements \cite{anandkumar2007type,mergen2006tbma, tbma_sayeed}. The outlined protocol can be considered as a form of joint source-channel coding, and is known as Type-Based Multiple Access (TBMA).

In this paper, we investigate the performance of TBMA in a multi-cell fog-based system, as seen in Fig. 1 and detailed in Sec. II and Sec. III. A key new aspect of this type of network deployments is that detection, via hypothesis testing, of the distributed QoIs can be either carried out locally at the ENs or centrally at the cloud. Edge detection is impaired by inter-cell interference, while cloud detection is subject to fronthaul capacity constraints. In contrast to recent works that considered distributed hypothesis testing over wireless channels \cite{distributed_hypothesis_1,distributed_hypothesis_2}, the goal in this paper is to detect the value of the QoI and not the joint distribution of all QoIs. The error exponent analysis presented in this paper (Sec. IV) provides insights into the performance comparison between edge and cloud processing, and the presented numerical results (Sec. V) validate our findings. Additional results can be found in the extended version of this paper \cite{rahif_information_centric}.


\par
\textbf{Notation:} Lower-case and upper case bold characters represent vectors and matrices respectively. $\mathbf{A}^{\mathsf{T}}$ denotes the transpose of matrix $\mathbf{A}$. $|\mathbf{A}|$ denotes the determinant of matrix $\mathbf{A}$. $A(i,j)$ denotes the element of $\mathbf{A}$ located at the $i$-th row and $j$-th column. $\mathcal{CN}(x|\mu,\sigma^2)$ is the probability density function (pdf) of a complex Gaussian random variable (RV) with mean $\mu$ and standard deviation $\sigma$. $C(f_1||f_2)$ represents the Chernoff information for the probability distributions $f_1$ and $f_2$. Given $a < b$, $[a,b]$ represents the segment of values between $a$ and $b$.$\langle a(t),b(t)\rangle = \int a(t)b(t) dt$ represents the correlation as applied to the given correlation interval.
\section{System and Signal Model}

\textit{System Model:} As illustrated in Fig.~\ref{fig:system_model}, we study a multi-cell fog-based IoT system that aims at detecting QoIs, such as pollution level, based on measurements received from IoT devices. There are $K$ cells, with a single-antenna EN and multiple IoT devices per cell. We assume that each QoI is described in each cell $c \in \{1, \ldots, K \}$ by a Random Variable (RV) $\theta^c$. RVs $\{\theta^c\}_{c=1}^{K}$ are arbitrarily correlated across cells, and each device in cell $c$ makes a noisy measurement of $\theta^c$. In this paper, we assume for simplicity of notation and analysis that each QoI can take two possible values, denoted as $\theta_0$ and $\theta_1$.

\par The IoT devices are interrogated periodically by their local EN during $L$ collection intervals, which are synchronized across all cells.
In each collection interval, a random number of devices in each cell $c$ transmit their measurements in the uplink using a grant-free access protocol based on TBMA \cite{anandkumar2007type}. Mathematically, in any collection interval $l=1,\ldots,L$, each IoT device in cell $c$ is active probabilistically, {\color{black}{independently of the observation being sensed}}, so that the total number $N_l^c$ of devices active in collection interval $l$ in cell $c$ is a Poisson RV with mean $\lambda$. All devices share the same spectrum and hence their transmissions generally interfere, both within the same cell and across different cells.
\par
We compare two different architectures to perform hypothesis testing in order to detect the QoIs: 
\textit{(i) Edge-based Hypothesis Testing (EHT):} Estimation of each QoI $\theta^c$ is done locally at the EN in cell $c$ based on the uplink signals received from the IoT devices, producing a local estimate $\hat{\theta}^c$ (see Fig.~\ref{fig:system_model}); and \textit{(ii) Cloud-based Hypothesis Testing (CHT):} The ENs are connected with orthogonal finite-capacity digital fronthaul links to a cloud processor with fronthaul capacity of $C\ \mathrm{[bit/s/Hz]}$. Each EN forwards the received signal upon quantization to the cloud processor using the fronthaul link. Unlike conventional C-RAN systems, here the goal is for the cloud to estimate all QoIs $\{ \theta^c \}_{c=1}^{K}$ (see Fig.~\ref{fig:system_model}).
\par
\textit{Signal Model: } When active, an IoT device $i$ in cell $c$ during the $l$-th collection makes a measurement $X_{i,l}^{c}$. We assume that the measurement takes values in an alphabet $\{1,2, \ldots, M \}$ of size $M$. 
The distribution of each observation $X_{i,l}^{c}$ depends on the underlying QoI as
\begin{equation}
\begin{aligned}
&\mathrm{Pr}[X^c_{i,l}=m|\theta^c=\theta_0]=p_0^c(m)\\
\mathrm{and }\ \ &\mathrm{Pr}[X^c_{i,l}=m|\theta^c=\theta_1]=p_1^c(m), \label{eq:dist}
\end{aligned}
\end{equation}
for $m=1,\ldots, M$.
In words, devices in cell $c$ make generally noisy measurements with $\theta^c$-dependent distributions $p^c_0(\cdot)$ and $p^c_1(\cdot)$. When conditioned on QoIs  $\{\theta^c\}$, measurements $X_{i,l}^{c}$ are i.i.d. across all values of the cell index $c$, device index $i$, and the collection index $l$.
\par We denote by $H_{i,l}^c \sim \cal{CN}$ $(\mu_{H}, \sigma^{2}_{H})$ the flat-fading Ricean fading channel, with mean $\mu_H$ and variance $\sigma^2_H$, from device $i$ to the $\mathrm{EN}$ in the same cell $c$ during collection interval $l$; and by $G_{i,l}^{c,c^\prime} \sim \mathcal{CN}(\mu_{G}, \sigma^{2}_{G})$, with mean $\mu_G$ and variance $\sigma^2_G$, the flat-fading Ricean fading channel from device $i$ in cell $c^\prime \neq c$ to the EN in cell $c$ during collection interval $l$. All channels are assumed i.i.d. across indices $i,l$ and $c$.
\section{Communication protocol and metrics}
\label{sec:protocol_metrics}
In this section, we detail the communication protocol and the performance metrics used.
\subsection{Communication Protocol}
Within the available bandwidth and time per-collection interval, as in \cite{mergen2006tbma}, we assume the presence of $M$ orthogonal waveforms $\{ \phi_m(t), m= 1,\ldots,M\}$, or preambles, with unit energy. According to TBMA, each waveform $\phi_m(t)$ encodes the value $m \in \{1,\ldots, M \}$ of the observations of a device. The signal transmitted by a device $i$ in cell $c$ that is active in interval $l$ is then given as $S_{i,l}^c(t) = \sqrt{E_s} \phi_{X_{i,l}^c}(t),$
that is, we have $S_{i,l}^c(t) = \sqrt{E_s} \phi_{m}(t)$ if the observed signal is $X_{i,l}^c(t)=m$, where $E_s$ is the transmission energy of a device per collection interval. Devices observing the same value $m$ hence transmit using the same waveform. As a result, the spectral resources required by TBMA scale with the number $M$ of observations values rather than with the total amount of packets sent by all the active devices, which may be much larger than $M$. \par 
The received signal at the $\mathrm{EN}$ in cell $c$ during the $l$-th collection can be written as
\begin{equation}
    Y_l^c(t) = \sum_{i=1}^{N_l^c} H_{i,l}^{c} S_{i,l}^c (t)+ \sum_{\substack{c^\prime=1 \\ c^\prime \neq c}}^{K}\sum_{i=1}^{N_l^{c^\prime}} G_{i,l}^{c,c^\prime} S_{i,l}^{c^\prime}(t) + W_l^c(t), \label{eq:ylc}
\end{equation}
where $W_l^c(t)\sim\mathcal{CN}(0,W_0)$ is white Gaussian noise, i.i.d. over $l$ and $c$, with power $W_0$. The first term in \eqref{eq:ylc} represents the contribution from the IoT devices in the same cell $c$, while the second term represents the contribution from devices from the remaining cells $c^\prime$. We emphasize that contributions related to the same preamble from different devices are not necessarily added coherently, but they only contribute to the average received energy for the preamble.  \par
Given the orthogonality of the waveforms $\{ \phi_m (t) \}_{m=1}^{M}$, a demodulator based on a bank of matched filters can be implemented at each EN without loss of optimality \cite{anandkumar2007type} (see \cite{JSC_MP_grant_free_IoT} for extensions). After matched filtering of the received signal with all waveforms $\{\phi_{m}(t)\}_{m=1}^{M}$ each EN $c$ obtains the $M \times 1$ vector
\begin{equation}
\begin{aligned}
      \mathbf{Y}_l^c &= \frac{1}{\sqrt{E_s}} [\langle \phi_1(t),Y_l^c(t)\rangle, \ldots , \langle \phi_M(t),Y_l^c(t)\rangle]^{\mathsf{T}} \\ 
      &= \sum_{i=1}^{N_l^c} H_{i,l}^{c} \mathbf{e}_{X_{i,l}^c}+ \sum_{\substack{c^\prime = 1 \\ c^\prime \neq c }}^{K}\sum_{i=1}^{N_l^{c^\prime}} G_{i,l}^{c, c^\prime} \mathbf{e}_{X_{i,l}^{c^\prime}} + \mathbf{W}_l, \label{eq:Yc}\\
\end{aligned}
\end{equation}
where $\mathbf{W}_l$ is a vector with i.i.d. $\mathcal{CN}(0, \mathrm{SNR}^{-1})$ elements, with $\mathrm{SNR}= E_s/W_0$; and $\mathbf{e}_m$ represents an $M \times 1$ unit vector with all zero entries except in position $m$.
\par
For detection of the QoIs, we study both EHT and CHT:
\par \textit{EHT:} Each EN $c$ produces an estimate $\hat{\theta^c}$ of the RV $\theta^c$ based on the received signals $\mathbf{Y}^c_l$ for all collection intervals $l=1,\ldots,L$, where $\mathbf{Y}^l_c$ is given in \eqref{eq:Yc}. \par
\textit{CHT:} Each EN $c$ compresses the received signals $\{ \mathbf{Y}^c_l \}^{L}_{l=1}$ across all $L$ collection intervals and sends the resulting compressed signals $\{ \hat{\mathbf{Y}}^c_l \}_{l=1}^{L}$ to the cloud. The cloud carries out joint detection of all QoIs $\{\theta^c \}_{c=1}^{K}$ producing estimates $\{ \Hat{\theta}^c\}_{c=1}^{K}$. 
\subsection{Performance Metrics} 
The performance of CHT and EHT will be evaluated in terms of the error exponent that describes the scaling of the joint error probability $\mathrm{P}_e$ as a function of the number $L$ of collections. The joint error probability is given by
\begin{equation}
\mathrm{P}_e = \mathrm{Pr}[\cup_{c=1}^{K} \{\hat{\theta^c} \neq \theta^c \}], \label{eq:joint_probability}    
\end{equation} 
where $\hat{\theta}^c$ is the estimate of the QoI $\theta^c$ obtained at $\mathrm{EN}$ $c$ or at the cloud, for EHT and CHT respectively. From large deviation theory, the detection error probability $P_e$ decays exponentially as \cite{cover2012elements}
\begin{equation}
    \mathrm{P}_e = \mathrm{exp}(-L E+ o(L))\ \ \ \mathrm{with}\ L\to\infty, \label{eq:Pe}
\end{equation}
where $o(L)/L \to 0$ as $L \to \infty$, for some error exponent $E$. We will hence be interested in the rest of this paper in computing analytically the error exponent $E$ for EHT and CHT.
\section{Asymptotic Performance}
\label{sec:asymptotic_performance}
In this section, we derive the error exponent $E$ in \eqref{eq:Pe} for the optimal detection when the number of collection intervals $L$ grows to infinity. In order to simplify the analysis, as in \cite{anandkumar2007type}, we will take the assumption of large average number of active devices, i.e., of large $\lambda$. This scenario is particularly relevant for mMTC \cite{ding20196g}.
\subsection{Edge-based Hypothesis Testing}
\label{sec:edge_detection_asymptotic}
With EHT, each $\mathrm{EN}$ in cell $c$ performs the binary test
\begin{equation}
    \mathcal{H}_0^c: \theta^c = \theta_0\   \mathrm{versus}\   \mathcal{H}_1^c:\theta^c = \theta_1 \label{eq:test_1}
\end{equation}
based on the available received signals $\mathbf{Y}^c=\{\mathbf{Y}_l^c \}_{l=1}^{L}$ in \eqref{eq:Yc}.
The optimum Bayesian decision rule that minimizes the probability of error at each EN chooses the hypothesis with the Maximum A Posteriori (MAP) probability. 
The error exponent $E$ in \eqref{eq:Pe} using EHT can be lower bounded as shown in the following proposition. \par
\textit{Proposition 1:} Under the optimal Bayesian detector, the error exponent $E$ in \eqref{eq:Pe} in the large-$\lambda$ regime and for any $0 < \rho < 1$ is lower bounded as $E \geq E^{edge} = \mathrm{min}_{c \in \{1,\ldots,K\}}E^c$, where
\begin{equation}
\begin{aligned}
    E^c  & = \min_{ \mathbf{k} \in \{ 0,1\}^{K-1}} \max_{\alpha \in [0,1]} \\ & \Bigg[\frac{1}{2} \sum_{m=1}^{M} \log\Big(\frac{\alpha \Sigma_{0,\mathbf{k}}^c(m,m) + (1-\alpha)\Sigma_{1,\mathbf{k}}^c(m,m)}{(\Sigma_{0,\textcolor{black}{\mathbf{k}}}^{c}(m,m))^{\alpha} (\Sigma_{1,\mathbf{k}}^{c}(m,m))^{\textcolor{black}{1-\alpha}}} \Big) \\& +\frac{\alpha(1- \alpha)}{2} \sum_{m=1}^{M} \frac{(\mu_{0,\mathbf{k}}^c(m) - \mu_{1,\mathbf{k}}^c (m))^2}{(\alpha \Sigma_{0,\mathbf{k}}^c(m,m) + (1-\alpha) \Sigma_{1,\mathbf{k}}^c (m,m) )} \Bigg] \label{eq:expo_edge}
\end{aligned}
\end{equation}
with $\mathbf{k} = \{k_{c^\prime}\}_{c^\prime = 1, c^\prime \neq c}^{K}$ \vspace{-5 mm}
\begin{equation}
   \mu_{k_c,\mathbf{k}}^c(m) = \mu_H \lambda p_{k_c}^c(m) + \mu_G \lambda  \sum_{c=1, c \neq c^\prime}^{K}p_{k_{c^\prime}}^{c^\prime}(m),
   \label{eq:dist_edge}
   \vspace{-5 mm}
\end{equation}
and
\vspace{-2 mm}
\begin{equation}
    \Sigma^c_{k_c,\mathbf{k}}(m,m)= \sigma^2_H \lambda p_{k_c}^c(m)  + \sigma^2_G  \lambda \sum_{c^\prime=1, c^\prime \neq c}^{K} p_{k_{c^\prime}}^c(m) + \frac{1}{\mathrm{SNR}}. \label{eq:BigSigma}
\end{equation}
\par \textit{Proof:} In a manner similar to \cite[Theorem 3]{anandkumar2007type}, the proof of the above theorem relies on the Central Limit Theorem (CLT) with random number of summands \cite[p. 369]{cover2012elements} and on the Chernoff Information \cite{cover2012elements}. We refer to the Appendix for more details. \qed
\par The term in \eqref{eq:expo_edge} being optimized over $\mathbf{k}$ corresponds to the Chernoff information \cite[Chapter 11]{cover2012elements} for the binary hypothesis test between the distributions of the received signal $\mathbf{Y}^c_l$ under hypotheses $\theta^c = \theta_0$ and $\theta^c = \theta_1$ when $\theta^{c^\prime}=\theta_{k_{c^\prime}}$. In fact, for large values of $\lambda$, when $\theta^c = \theta_{k_c}$ and $\theta^{c^\prime} = \theta_{k_{c^\prime}}$, the received signal $\mathbf{Y}^c_l$ in \eqref{eq:Yc} can be shown to be approximately distributed as $\mathcal{CN}(\boldsymbol{\mu}_{k_c,\mathbf{k}}^c,\mathbf{\Sigma}_{k_c,\mathbf{k}}^c)$, with mean vector $\boldsymbol{\mu}_{k_c,\mathbf{k}}^c = [\mu_{k_c,\mathbf{k}}^c(1) , \ldots , \mu_{k_c,\mathbf{k}}^c(M) ]^{\mathsf{T}}$ and diagonal covariance matrix $\mathbf{\Sigma}_{k_c,\mathbf{k}}^c$ with diagonal elements $\Sigma_{k_c,\mathbf{k}}^c (m,m)$.
\subsection{Cloud-based Hypothesis Testing}
\label{sec:cloud_detection}
The cloud tackles the $2^K$-ary hypothesis testing problem of distinguishing among hypotheses $\mathcal{H}_{k_1, \ldots, k_K}:(\theta^1,\ldots,\theta^K) = (\theta_{k_1},\ldots,\theta_{k_K})$ for $k_c \in \{0,1 \}$ on the basis of the quantized signals $\{ \hat{\mathbf{Y}}_l \}^{L}_{l=1}$ received from both ENs on the fronthaul links.\par
Following a standard approach, see, e.g., \cite{bookcransimeone}, the impact of fronthaul quantization is modeled as an additional quantization noise. In particular, the signal received at the cloud from EN $c$ can be written accordingly as $\hat{\mathbf{Y}}_{l}^c = \mathbf{Y}^{c}_{l} + \mathbf{Q}^c_l$, where $\mathbf{Q}^c_l$ represents the quantization noise vector. As in most prior references (see, e.g., \cite{bookcransimeone}), the quantization noise vector $\mathbf{Q}^c_l$ is assumed to have i.i.d. elements being normally distributed with zero mean and variance $\sigma^2_{q^c}$. Furthermore, from rate-distortion theory, the fronthaul capacity constraint implies the following inequality, for each EN $c$
\begin{equation}
    M C \geq I(\mathbf{Y}^c_{l} ; \hat{\mathbf{Y}}^c_{l}). \label{eq:capacity_contraint}
\end{equation}
This is because the number of bits available to transmit each measurement $\hat{\mathbf{Y}}^c_l$ is given by $C$ bits per symbol, or, equivalently, per orthogonal spectral resource; that is, $MC$ bits in total for all $M$ resources. From \eqref{eq:capacity_contraint}, one can in principle derive the quantization noise power $\sigma^2_{q^c}$. However, evaluating the mutual information in \eqref{eq:capacity_contraint} directly is difficult due to the non-Gaussianity of the received signals $\mathbf{Y}_l^c $. To tackle this issue, we bound the mutual information term in \eqref{eq:capacity_contraint} using the property that the Gaussian distribution maximizes the differential entropy under covariance constraints \cite{cover2012elements}, obtaining the following Lemma. In what follows, we denote $\mathbf{k}=\{k_c\}_{c=1}^{K}$.\par
\textit{Lemma 1:} The quantization noise power can be upper bounded as $\sigma^2_{q^c} \leq \Bar{\sigma}^2_{q^c}$, where $\Bar{\sigma}^2_{q^c}$ is obtained by solving the non-linear equation
\begin{equation}
\begin{aligned}
    & MC  = \frac{1}{2} \sum_{m=1}^{M} \log  \\  & \Bigg( \frac{ \sum_{\mathbf{k} \in \{0,1\}^K} \mathrm{Pr}( \theta^1 = \theta_{k_1},\ldots,\theta^K = \theta_{k_K}) \Sigma^{c}_{\mathbf{k}} (m,m) + \sigma_{q^c}^2}{(\sigma_{q^c}^2)^M} \Bigg). \label{eq:tosolve_quantization}
    \end{aligned}
\end{equation}
with $\Sigma^{c}_{\mathbf{k}}$ given in equation \eqref{eq:BigSigma}.
\par \textit{Proof:} See \cite[Appendix A]{rahif_information_centric} for details. \qed \\
\textit{Proposition 2:} Under optimal detection, the error exponent $E$ in \eqref{eq:Pe} in the large-$\lambda$ regime for CHT can be lower bounded as $E \geq E^{cloud} = \mathrm{min}_{\mathbf{k} \in \{0,1 \}^K} E_{\mathbf{k}}$, where
\begin{equation}
\begin{aligned}
    E_{\mathbf{k}} &=\min_{ \mathbf{k}^\prime \neq \mathbf{k}} \max_{\alpha \in [0,1]} \Big[ \frac{1}{2} \mathrm{log} \frac{|\alpha \mathbf{\Sigma}_{\mathbf{k}} + (1-\alpha)\mathbf{\Sigma}_{\mathbf{k}^\prime}|}{|\mathbf{\Sigma}_{\mathbf{k}}|^\alpha|\mathbf{\Sigma}_{\mathbf{k}^\prime}|^{1-\alpha}} \\&+ \frac{\alpha (1-\alpha)}{2}(\boldsymbol{\mu}_{\mathbf{k}} - \boldsymbol{\mu}_{\mathbf{k}^\prime})^{\mathsf{T}}(\alpha \mathbf{\Sigma}_{\mathbf{k}} + (1-\alpha)\mathbf{\Sigma}_{\mathbf{k}^\prime})^{-1} \\ & \times (\boldsymbol{\mu}_{\mathbf{k}} - \boldsymbol{\mu}_{ \mathbf{k}^\prime}) \Big], \label{eq:expo_cloud}
    \end{aligned}
\end{equation}
where the entries of the $KM \times 1$ vector $\boldsymbol{\mu}_{\mathbf{k}}$ are defined as
\begin{equation}
    \mu_{\mathbf{k}}(m^\prime) = \mu_{\mathbf{k}}^c(m)\ \ \mathrm{for}\ m^\prime=(c-1)M,\ldots,cM
    \label{eq:cloud_mu}
\end{equation}
with $\mu_{\mathbf{k}}^c(m)$ defined in \eqref{eq:dist_edge}, and the entries of the $2M \times 2M$ covariance matrix $\mathbf{\Sigma}_{\mathbf{k}}$ given as
\begin{equation}
\begin{aligned}
    &\Sigma_{\mathbf{k}} (m^{\prime},m^{\prime}) = \Sigma_{\mathbf{k}}^1 (m,m) + \sigma^2_{q^c} \ \ \mathrm{for }\ m^{\prime}=(c-1)M,\ldots,cM,\\
    & \Sigma_{\mathbf{k}} (m^\prime,M+m^\prime) = \Sigma_{\mathbf{k}} (M+m^\prime,m^\prime) = \\& p_{k_c}^{c}(m)(1-p_{k_c}^{c}(m)) \lambda \mu_H \mu_G + p_{k}^{c^\prime}(m)(1 - p_{k}^{c^\prime}(m))\lambda \mu_H \mu_G\ \ \\ & \mathrm{for }\ m=1,\ldots,M,\ \mathrm{and}\ m^{\prime}=(c-1)M,\ldots,cM, \label{eq:param_cloud_2}
    \end{aligned}
\end{equation}
where $\Sigma^c_{\mathbf{k}}(m,m)$ is defined in \eqref{eq:BigSigma} and all other entries of matrix $\mathbf{\Sigma}_{\mathbf{k}}$ are zero.\par
\textit{Proof:} The proof follows in a manner similar to \textit{Proposition 1} and uses Sanov's Theorem \cite[p. 362]{cover2012elements} as detailed in \cite[Appendix C]{rahif_information_centric}. \qed 
\par 
The term in \eqref{eq:expo_cloud} being optimized over $ \mathbf{k}^{\prime} \in \{0,1\}^K$ corresponds to the Chernoff information for the binary test between the distribution of the signal received at the cloud under hypotheses $\mathcal{H}_{\mathbf{k}}:(\theta^1 = \theta_{k_1},\ldots,\theta^{K} = \theta_{k_K})$ and $\mathcal{H}_{\mathbf{k^\prime}}:(\theta^1 = \theta_{k_1^\prime},\ldots,\theta^{K} = \theta_{k_K^\prime})$. As discussed above, for large $\lambda$, the signal received at the cloud under hypothesis $\mathcal{H}_{\mathbf{k}}$ is approximately distributed as $\mathcal{CN}(\boldsymbol{\mu}_{\mathbf{k}},\mathbf{\Sigma}_{\mathbf{k}})$, where the elements of the mean vector $\boldsymbol{\mu}_{\mathbf{k}}$ and covariance matrix $\mathbf{\Sigma}_{\mathbf{k}}$ are described in \eqref{eq:cloud_mu} and \eqref{eq:param_cloud_2}.
\textcolor{black}{
\subsection{Edge vs Cloud-Based Hypothesis Testing}
\label{sec:edge_vs_cloud}
In this section, we prove that the performance of CHT is superior to EHT as long as the inter-cell channel power gain power $\sigma^2_G$ is sufficiently large. The main result can be summarized in the following theorem.}
\par \textcolor{black}{\textit{Theorem 1: } The error exponents derived in \textit{Proposition 1} and \textit{Proposition 2} satisfy the following limits
\begin{equation}
    \lim_{\sigma^2_G \to \infty} E^{edge} = 0\ \ \mathrm{and}\ \ \lim_{\sigma^2_G \to \infty} E^{cloud} > 0. \label{eq:theorem_1}
\end{equation}}
\par \textcolor{black}{\textit{Proof: }The proof can be found in \cite[Appendix D]{rahif_information_centric}}\qed \par
\textcolor{black}{\textit{Theorem 1} implies that, for high inter-cell power gains, EHT leads to vanishing error exponent, while this is not the case for CHT. This demonstrates that the performance of EHT is inter-cell interference limited, while this is not the case for CHT. In practice, as shown via numerical results in Sec. \ref{sec:numerical_results}, fairly low interference levels are sufficient for CHT to outperform EHT.}
\begin{figure}[t]
	\centering
	\includegraphics[height= 5.3 cm, width= 8.5 cm]{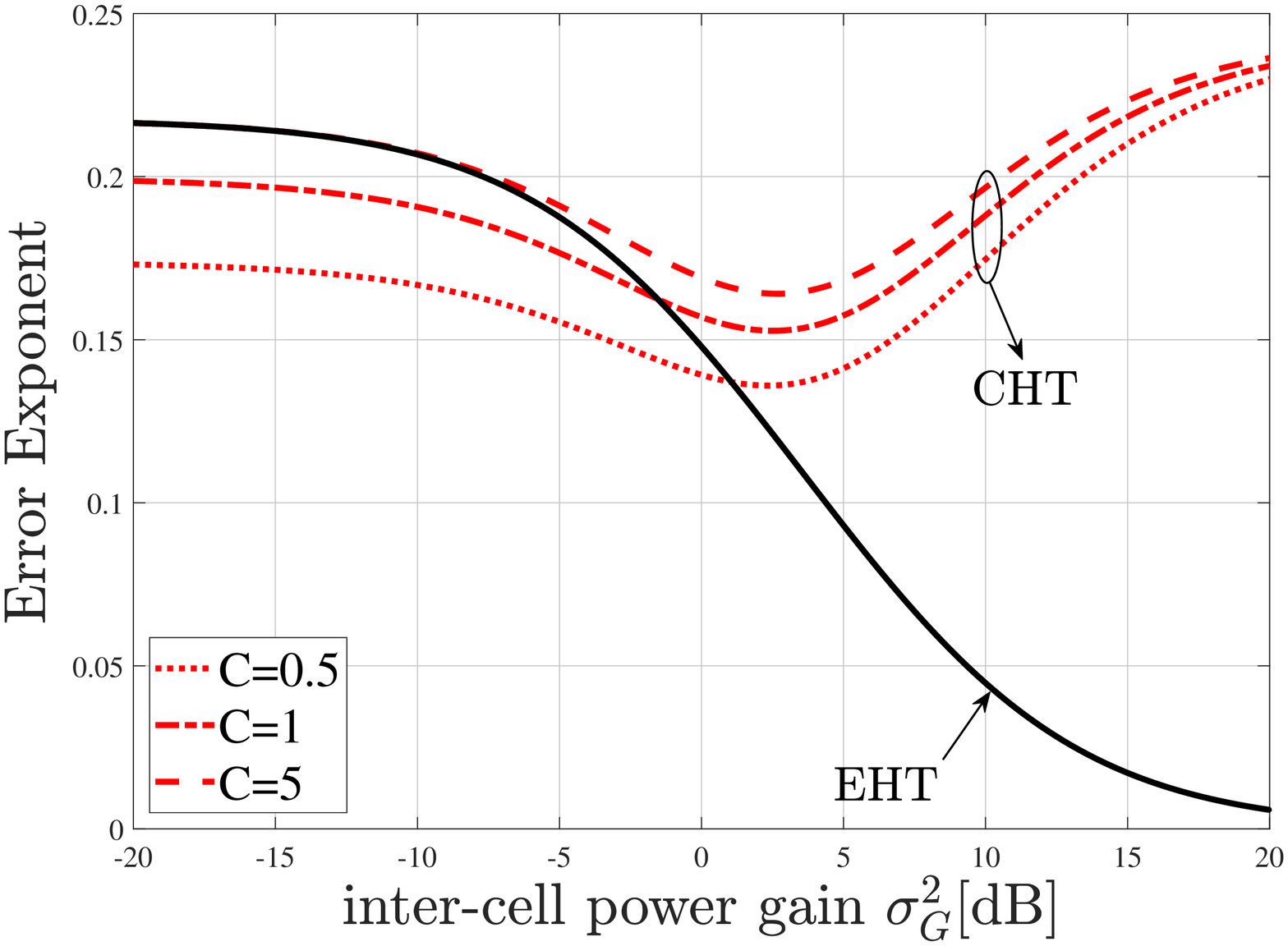}
	\caption{Error exponent for EHT and CHT as function of the inter-cell power gain $\sigma^2_G$ ($\mu_H=1,\sigma^2_H=1,\mu_G=0$, $\lambda=4$, and $\mathrm{SNR}= -1\ \mathrm{dB}$).}
	\label{fig:expo_sigma2G} \vspace{-5 mm}
\end{figure}
\section{Numerical Results}
\label{sec:numerical_results}
In this section, we provide numerical simulations to evaluate the performance of both CHT and EHT. Unless specified otherwise, we fix the following values for the parameters $\mu_H=1,\sigma^2_H=1,\mu_G=0$, $\lambda=4$ and $K=2$ cells. The joint distribution of QoIs is defined as
\begin{equation}
  f (\theta^c , \theta^{c^\prime}) = \frac{\rho}{2} 1_{\{ \theta^c = \theta^{c^\prime}\}} + \frac{1-\rho}{2} 1_{\{\theta^c \neq \theta^{c^\prime}\}}   \label{eq:param_dist}
\end{equation}
where $ 0 \leq \rho \leq 1$ represents a ``correlation" parameter that measures the probability that the two QoIs have the same value, i.e., $\rho = \mathrm{Pr} [\theta^c = \theta^{c^\prime}]$.
Note that under \eqref{eq:param_dist}, both values of the QoI are equiprobable, i.e., $\mathrm{Pr}(\theta^c = \theta_j)=0.5$ for $j \in \{0,1\}$ and $c \in \{1, 2\}$. Furthermore, when $\rho=0.5$, the two QoIs are independent.
\par
In Fig. \ref{fig:expo_sigma2G}, we plot the error exponent for both EHT and CHT with different values of $C$ as function of the inter-cell power gain $\sigma^2_G$. As $\sigma^2_G$ increases, the performance of edge detection is seen to decrease, since interference from the other cell is treated as noise by the edge. In contrast, inline with the theoretical results in \textit{Theorem 1}, CHT is able to benefit from a sufficiently large inter-cell interference due to centralized processing. We note that, the same U-shaped behavior is observed for the uplink throughput in C-RAN as function of the inter-cell interference \cite{simeone2012cooperative}.  Furthermore, a larger fronthaul capacity leads to an improved detection performance, since measurements are received at the cloud with a better resolution. \par
\begin{figure}[h]
	\centering
	\includegraphics[height= 5.3 cm, width= 8.5 cm]{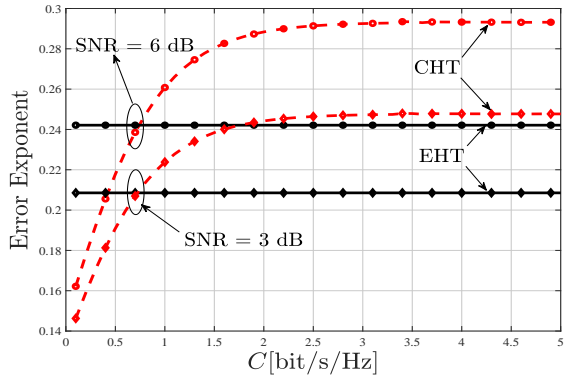}
	\caption{Error exponent for EHT and CHT as function of the fronthaul capacity $C$ ($\mu_H=1,\sigma^2_H=1,\mu_G=0, \sigma^2_G=0$, and $\lambda=4$).}
	\label{fig:expo_C} \vspace{-8 mm}
\end{figure}

In Fig. \ref{fig:expo_C}, we plot the error exponent as function of the fronthaul capacity $C$. For low values of $C$, EHT outperforms CHT since in this regime, the quantization noise is large and thus measurements are received with low resolution. In contrast, CHT outperforms EHT for high enough values of $C$.
\section{Conclusions}
\label{sec:conclusions}
This paper considers the problem of detection of Quantities of Interest (QoIs) at the edge or the cloud of a fog-based IoT network. The performance of cloud-based detection was demonstrated analytically and via numerical results to be superior to edge-based detection for sufficiently high fronthaul capacity and inter-cell interference. As for future research directions, we mention the study of the coexistence of heterogeneous IoT services with different service requirements.
\section*{Appendix: Proof of Proposition 1}
\label{sec:appendix_a}
From the union bound $P_e \leq \sum_{c=1}^{K} P_e^c$ with $P_e^c = \mathrm{Pr}[\hat{\theta}^c \neq \theta^c]$ and the identity $P_e^c = \frac{1}{2^{K-1}} \sum_{\mathbf{k}\in \{0,1\}^{K-1}} \mathrm{Pr}[\hat{\theta}^c \neq \theta^c | \{\theta^{c^\prime} = \theta_{k_c^\prime},\ \mathrm{for}\ c^\prime \neq c \in \{1,\ldots,K\}\}]$, we directly obtain the lower bound on the error exponent
\begin{equation}
     E \geq   \min_{c \in \{1,\ldots,K \}} \min_{ \mathbf{k} \in \{0,1 \}^{K-1}} E^c , \label{eq:E_edge_bound}
\end{equation}
where $E^c=- \lim_{L \to \infty} \frac{1}{L} \log \mathrm{Pr}[\hat{\theta}^c \neq \theta^c |\{\theta^{c^\prime} = \theta_{k_c^\prime},\ \mathrm{for}\ c^\prime \neq c \in \{1,\ldots,K\}\}]$ is the error exponent for detection of QoI $\theta^c$ conditioned on the condition $\theta^{c^\prime}=\theta_{k_{c^\prime}}$ for $c^\prime \neq c$.
Under optimal Bayesian detection, the error exponent $E^c_k$ is given by the Chernoff information \cite[Chapter 11]{cover2012elements} as 
\begin{equation}
    E^c = C(f_{0,\mathbf{k}}(\mathbf{Y}^c_l),f_{1,\mathbf{k}}(\mathbf{Y}^c_l)), \label{eq:E_edge_chernoff}
\end{equation}
where we have denoted $f_{k_c,\mathbf{k}}(\mathbf{Y}^c_l) = f(\mathbf{Y}^c_l|\theta^c = \theta_j , \{\theta^{c^\prime} = \theta_{k_c^\prime},\ \text{for}\ c^\prime \neq c \in \{1,\ldots,K\}\})$ for brevity. Computing the error exponent in \eqref{eq:E_edge_chernoff} requires finding the distributions $f_{k_c,\mathbf{k}}(\mathbf{Y}^c_l)$. Following \cite{anandkumar2007type}, this can be approximated by a Gaussian distribution in the regime of large $\lambda$ thanks to the Central Limit Theorem (CLT) with random number of summands \cite[p. 369]{billingsley2008probability}.
In particular, referring to \cite{anandkumar2007type} for details, we can conclude that, when $\lambda \to \infty$, the conditional distribution $f_{k_c,\mathbf{k}}(\mathbf{Y}^c)$ tends in distribution to $\mathcal{CN}(\boldsymbol{\mu}_{k_c,\mathbf{k}} , \mathbf{\Sigma}_{k_c,\mathbf{k}})$, where $\boldsymbol{\mu}_{k_c,\mathbf{k}}$ and $\mathbf{\Sigma}_{k_c,\mathbf{k}}$ are the mean vector and covariance matrix respectively when $\theta^c = \theta_{k_c}$ and $\theta^{c^\prime}=\theta_{k_{c^\prime}}$ and are defined in \eqref{eq:dist_edge} and \eqref{eq:BigSigma}.
\par The Chernoff Information between two Gaussian distributions can be obtained by maximizing over $\alpha \in [0,1]$ the $\alpha$-Chernoff information defined as \cite{nielsen2011chernoff}
\begin{equation}
\begin{aligned}
    & C_{\alpha}(f_{0,\mathbf{k}}({\mathbf{Y}}^c_l),f_{1,\mathbf{k}}({\mathbf{Y}}^c_l)) =\\ &\frac{1}{2} \mathrm{log} \frac{|\alpha \mathbf{\Sigma}_{0,\mathbf{k}} + (1-\alpha)\mathbf{\Sigma}_{1,\mathbf{k}}|}{|\mathbf{\Sigma}_{0,\mathbf{k}}|^\alpha|\mathbf{\Sigma}_{1,\mathbf{k}}|^{1-\alpha}} + \frac{\alpha (1-\alpha)}{2}(\boldsymbol{\mu}_{0,\mathbf{k}} - \boldsymbol{\mu}_{1,\mathbf{k}})^{\mathsf{T}}\\ & \times (\alpha \mathbf{\Sigma}_{0,\mathbf{k}} + (1-\alpha)\mathbf{\Sigma}_{1,\mathbf{k}})^{-1} (\boldsymbol{\mu}_{0,\mathbf{k}} - \boldsymbol{\mu}_{1,\mathbf{k}}). \label{eq:alpha_chernoff}
\end{aligned}
\end{equation} 
By plugging in \eqref{eq:E_edge_bound} and \eqref{eq:alpha_chernoff} the expressions of $\boldsymbol{\mu}_{k_c,\mathbf{k}}$ and $\mathbf{\Sigma}_{k_c,\mathbf{k}}$ and using \eqref{eq:E_edge_chernoff} we obtain the desired result.\qed

\bibliographystyle{IEEEtran}
\bibliography{Biblio}
\end{document}